\pdfoutput=1
\documentclass[11pt, a4paper]{article}

\usepackage{amsthm}
\usepackage{amsmath}
\usepackage{amssymb}
\usepackage{amsfonts}
\usepackage{latexsym}
\usepackage{booktabs}
\usepackage{multirow}
\usepackage{graphicx}
\usepackage{natbib}
\usepackage{pdfpages}
\usepackage{algorithm}
\usepackage{algpseudocode}
\usepackage{parskip}
\usepackage[left = 2.5cm, right = 2.5cm, top = 2.5cm, bottom = 3cm]{geometry}
\usepackage[colorlinks,citecolor=black,urlcolor=black]{hyperref}

\newtheoremstyle{example}
{3pt} 
{3pt} 
{} 
{0\parindent} 
{\bf}
{:} 
{.5em} 
{} 
\newtheoremstyle{theorem}
{3pt} 
{3pt} 
{\em} 
{0\parindent} 
{\bf}
{:} 
{.5em} 
{} 
\theoremstyle{example} 
\theoremstyle{theorem} 
\theoremstyle{lemma} 
\theoremstyle{corollary} 

\begin{document}

\title{Davidson-Luce model for multi-item choice with ties\thanks{The work of David Firth and Ioannis Kosmidis was
  supported by the Alan Turing Institute under EPSRC grant
  EP/N510129/1.}}

\author{ David Firth\thanks{
           Department of Statistics, University of Warwick, UK; and
           The Alan Turing Institute, London, UK.
           ORCiD: 0000-0003-0302-2312.
           \href{mailto:d.firth@warwick.ac.uk}{d.firth@warwick.ac.uk}}
           \and
           Ioannis Kosmidis\thanks{
           Department of Statistics, University of Warwick, UK; and
           The Alan Turing Institute, London, UK.
           ORCiD: 0000-0003-1556-0302.}
           \and
           Heather L. Turner\thanks{
           Department of Statistics, University of Warwick, UK.
           ORCiD: 0000-0002-1256-3375.}
           }
           
\maketitle

\begin{abstract}
  This paper introduces a natural extension of the pair-comparison-with-ties model of Davidson (1970, J.~Amer.~Statist.~Assoc), to allow for ties when more than two items are compared.  
  Properties of the new model are discussed.
  It is found that this `Davidson-Luce' model retains the many appealing features of Davidson's solution, while extending the scope of application substantially beyond the domain of pair-comparison data.
  The model introduced here already underpins the handling of tied rankings in the \textbf{PlackettLuce} \emph{R} package. \\

  
\end{abstract}

\section{Background: Pair comparisons}
\label{intro}

\subsection{Bradley-Terry model and Davidson's generalization for ties}
\label{bradley-terry}

A commonly used statistical model for pair-comparison data is the so-called Bradley-Terry model \citep{Bradley1952}, in which a binary outcome `$i$ is preferred to $j$' or `$i$ beats $j$' is assumed to have probability in the form
\[
\frac{\alpha_i}{\alpha_i + \alpha_j}.
\]
In the Bradley-Terry model each `item' (or `player') $i$ has their own unobserved `strength' or `ability' $\alpha_i > 0$, and it is the relative values of $\alpha_i$ and $\alpha_j$ that determine the win-probabilities when $i$ and $j$ are compared.

The Bradley-Terry model is a logit-linear model for the binary outcome ($i$ wins, or $j$ wins); and the ratio $\alpha_i / \alpha_j$ is readily interpretable as the \emph{odds} (on $i$ winning, in a contest between $i$ and $j$).
The Bradley-Terry model has also been shown \citep{Luce1959, Luce1977} to follow from a simple and appealing axiom for behaviour when making choices among items.
For choosing the preferred item from a finite set $S$, Luce's axiom implies choice probabilities
\[
\frac{\alpha_i}{\sum_{k \in S} \alpha_k}\quad (i \in S),
\]
from which the Bradley-Terry model follows whenever $S$ contains only two elements.

The Bradley-Terry model's outcomes are strictly binary: ties are not permitted.
\citet{Davidson1970} shows how to generalize the Bradley-Terry model to accommodate ties, in a way that does not violate Luce's axiom.
The Davidson model stipulates, for the three possible outcomes \{$i$ wins, $j$ wins, tie\} in a comparison of items $i$ and $j$, probabilities (summing to 1) as follows:
\begin{center}
\begin{tabular}{lccc}
\textbf{Outcome}:\qquad & $i$ wins & $j$ wins & tie \\
\textbf{Probability} (proportional to):\qquad & $\alpha_i$ & $\alpha_j$ &
$\delta(\alpha_i\alpha_j)^{1/2}$ \\
\end{tabular}
\end{center}
The Davidson model thus incorporates a single additional parameter, $\delta$, which describes the prevalence of ties; different values of $\delta$ will be appropriate in different application contexts.

\subsection{Properties of the Davidson model}

Some well-known properties of the Davidson model are as follows:

\begin{enumerate}
\def\labelenumi{\arabic{enumi}.}
\item
 The geometric mean $(\alpha_i \alpha_j)^{1/2}$ has the same dimension as $\alpha_i$ and $\alpha_j$; that is to say, their units of measurement are the same.  This makes the tie-prevalence parameter $\delta$ dimensionless, and straightforwardly interpretable.  Specifically, the probability of a tie in any comparison between items of equal strength (i.e., $\alpha_i = \alpha_j$) is ${\delta}/({2 + \delta})$. 
\item
  Conditional upon the outcome \emph{not} being a tie, the probability that $i$ wins is $\alpha_i/(\alpha_i + \alpha_j)$, exactly as in the Bradley-Terry model for binary outcomes.
  In this way the Davidson model maintains compatibility with Luce's axiom.
\item
  Like the Bradley-Terry model, the Davidson generalization depends on the strengths only through their \emph{relative} values.
  The scale --- or unit of measurement --- of strengths $\{\alpha_i, \alpha_j,\ldots\}$ is immaterial.
\item
  For any fixed value of $\delta$, the tie probability is proportional to $(\alpha_i \alpha_j)^{1/2}/(\alpha_i + \alpha_j)$ and is maximized when $\alpha_i=\alpha_j$.
  That is, ties are most likely when the items being compared have equal strength.
\item
  The Davidson model is a full exponential family model, and so maximum likelihood estimation of the parameters (the strengths $\{\alpha_i, \alpha_j, \ldots\}$ and the tie prevalence $\delta$) simply equates sufficient statistics with their expectations under the model.
  The sufficient statistics are
  \begin{itemize}
  \item
    for each item, its observed number of `wins' plus half its observed number of ties;
  \item
    the total number of ties seen, in all comparisons made.
  \end{itemize}
See, e.g., \citet{Fienberg1979} for full details of the model's representation in log-linear form, and consequent solution of the likelihood equations in standard software.
\item
  The preceding property has a neat implication when the Davidson model is applied to a `balanced round-robin' tournament among $n$ items, where every item is compared with every other item the same number of times.
  In that context the maximum likelihood estimates $\{\hat\alpha_i:\, i=1,\ldots,n\}$ are ordered in exactly the same way as would be simple, item-specific `points totals', with 2 points awarded for a win and 1 point for a tie \citep{Davidson1970}.
  This holds regardless of the value of $\delta$.
\end{enumerate}

\section{More than two items: Davidson-Luce model}

\subsection{Preamble}
\label{preamble}

In this section we extend the Davidson model to comparisons involving more than two items.  The new `Davidson-Luce model' is designed to retain 
the key properties of the Davidson model.

In general we will suppose that a choice is to be made (i.e., a winner is to be determined) from $r$ items. 
The outcome can be a single `best' item, or a tie between two or more of the items under comparison.

The model is introduced first for $r=3$, before giving its general definition for any finite $r$.

\subsection{Choice from three items}

\label{three_items}

With three items $i, j, k$, there are 7 possible outcomes.
We will label these here as

\begin{itemize}
\item
  $i$, $j$, $k$ (a single item wins)
\item
  $ij$, $jk$, $ik$ (two items are tied winners)
\item
  $ijk$ (all three items are tied winners)
\end{itemize}

The Davidson-Luce model in this case specifies 7 probabilities that sum to 1, in the following proportions:

\begin{center}\small
\begin{tabular}{lccccccc}
\textbf{Outcome}:\qquad & $i$ & $j$ & $k$ & $ij$ & $jk$ & $ik$ & $ijk$ \\
\textbf{Probability} (proportional to):\qquad & $\alpha_i$ & $\alpha_j$ & $\alpha_k$ &
$\delta_2(\alpha_i\alpha_j)^{1/2}$ &
$\delta_2(\alpha_j\alpha_k)^{1/2}$ &
$\delta_2(\alpha_i\alpha_k)^{1/2}$ &
$\delta_3(\alpha_i\alpha_j\alpha_k)^{1/3}$ \\
\end{tabular}
\end{center}

In this model there are two separate tie-prevalence parameters, $\delta_2\ge 0$ and $\delta_3\ge 0$, for the prevalence of 2-way ties and 3-way ties respectively.
The interpretation of strengths $\alpha_i, \alpha_j, \alpha_k$ is still as in the Luce model: conditional upon the outcome being an outright win for one item, the probabilities are in the ratios $\alpha_i:\alpha_j:\alpha_k$.

Still it is the case --- as in the Bradley-Terry, Luce and Davidson models --- that only \emph{relative} values of the strength parameters affect the model.
Moreover, as before, the tie probabilities are all maximized when strengths are equal.

The interpretation of $\delta_2$ is like that of $\delta$ in the Davidson model.
For example, conditional upon $k$ not being included in the winning choice, the possible outcomes are $\{i, j, ij\}$, in which case $\delta_2/(2 + \delta_2)$ is --- as before --- the probability of a 2-way tie between $i$ and $j$ when $\alpha_i = \alpha_j$.
Alternatively, if we condition only upon the outcome not being a 3-way tie, then with $\alpha_i=\alpha_j=\alpha_k$ the probability of a 2-way tie is $\delta_2/(1 + \delta_2)$.

The interpretation of $\delta_3$, similarly, is simplest in terms of the hypothetical situation of equal strengths $\alpha_1=\alpha_2=\alpha_3$ (i.e., the situation where, for any given value of $\delta_3>0$, the probability of a 3-way tie is maximized).
The probability of a 3-way tie is then $\delta_3/(3 + 3\delta_2 + \delta_3)$.

The extensions of properties 5 and 6 listed above for the Davidson model are as follows.
The model is a full exponential family, whose sufficient statistics are:

\begin{itemize}
\item
  for each item, its observed number of outright `wins', plus $\frac{1}{2}$ of its observed number of 2-way ties, plus $\frac{1}{3}$ of its observed number of 3-way ties;
\item
  the total number of 2-way ties seen, in all comparisons made;
\item
  the total number of 3-way ties seen, in all comparisons made.
\end{itemize}

As a consequence, in a balanced round-robin tournament of 3-way comparisons involving $n$ items in total, the maximum likelihood estimates $\hat\alpha_i,\ldots,\hat\alpha_n$ are ordered in exactly the same way as are simple, item-specific `points totals', with 6 points awarded for an outright win, 3 points for a 2-way tied win, and 2 points for a 3-way tie.
This holds regardless of the values of $\delta_2$ and $\delta_3$.

Further discussion of the properties of the Davidson-Luce model 
is deferred to section \ref{properties}.
In the next subsection we show how this Davidson-Luce model extends, in an obvious way, to a choice made from any number $r$ of items.

\subsection{Choice from any finite set}

\label{general_r}

The model for $r=3$, as described above, immediately suggests the form of the Davidson-Luce model for any $r$.

In any given comparison, label the items being compared by $\{i_1,i_2,\ldots,i_r\}$, 
and denote by $T$  the set of possible `winning' choices that might be made from the $r$ items being compared. For example, $T = \{i_2\}$ indicates an outright winner, $T = \{i_1, i_2\}$ indicates a 2-way tie, and so on, up to and including the possibility $T = \{i_1,i_2,\ldots,i_r\}$, which indicates that all $r$ items tied.

The Davidson-Luce model stipulates that the probability of any such choice $T$ is proportional to
\begin{equation}
p_T = \delta_t\left(\prod_{i \in T} \alpha_i\right)^{1/t},
\end{equation}
where $t$ denotes the cardinality of set $T$.
Thus $t$ can take values in $\{1,\ldots,r\}$. The adjustable tie-prevalence parameters are $\delta_2,\ldots, \delta_r$; the value of $\delta_1$ can be set arbitrarily to be 1, so $\delta_1$ is not actually a parameter in the model but is included here for presentational tidiness.

The constant of proportionality is just the normalizing constant, the reciprocal of the sum of $p_T$ over all possible choice sets $T$.
That normalizing constant can be  straightforwardly computed, if needed, but, it involves a rapidly increasing number of terms as the value of $r$ increases. 
The model's log-linear representation, which follows as a direct extension of \citet{Fienberg1979}, allows for simple iterative computation of estimates and associated standard errors without any need to evaluate the likelihood itself. 
A numerical illustration is provided in the Appendix, to show how this works in detail.

\hypertarget{basic-properties-of-the-model}{%
\subsection{Basic properties of the model}\label{basic-properties-of-the-model}}

\label{properties}

The \citet{Davidson1970} model is a special case of the Davidson-Luce model,
with $r \equiv |S| = 2$ and $\delta_2 \equiv \delta$. The Luce model \citep{Luce1959, Luce1977} is
the special case in which ties are not allowed: that is, $\delta_t \equiv 0$ for all
$t>1$.

Here we briefly describe how the Davidson model
properties listed above (in section \ref{intro})
extend to the Davidson-Luce model.

\begin{enumerate}
\def\labelenumi{\arabic{enumi}.}
\item
  The geometric means $\left(\prod_{i \in T} \alpha_i\right)^{1/t}$ all have the same dimensions as the strengths $\{\alpha_i\}$, and so the tie-prevalence parameters $\delta_2,\ldots,\delta_r$ are all dimensionless.
  It was shown in section \ref{three_items} above how to construct meaningful interpretations for those parameters.
\item
  Conditional upon the outcome of a comparison \emph{not} being a tie, the probability that $i$ wins is $\alpha_i/\sum_{k \in S} \alpha_k$, for any $i$ in the comparison set $S$.
  The Davidson-Luce model thus maintains compatibility with Luce's axiom.
\item
  As before, dependence on item strengths is only ever through their \emph{relative} values.
\item
  The tie probabilities all are in the form of geometric means, which are maximized when the items being compared have equal strengths.
\item
  The Davidson-Luce model is still a full exponential family model as before, the sufficient statistics being
  
  \begin{itemize}
  \item
    for each item, the total number of wins, counting a tied win fractionally in the obvious way;
  \item
    the total numbers of ties seen, of each order (i.e., the count of 2-way ties,
    the count of 3-way ties, etc.).
  \end{itemize}
A straightforward extension of the log-linear representation in \citet{Fienberg1979} leads to efficient solution of maximum likelihood equations --- without any need to compute the likelihood itself --- using standard software for generalized linear models.
\item
  As already exemplified in section \ref{three_items}, the Davidson-Luce model continues to yield exact agreement with points-based league tables for fully balanced tournaments, provided that points are divided equally whenever items share a tied win.
\end{enumerate}

In summary, then: the Davidson-Luce model retains the many appealing features of the Davidson model for ties, while extending the scope of application substantially beyond the limited domain of pair-comparison data.

\section{Concluding remarks}

\label{remarks}

A specific application of the ideas developed here is to the Plackett-Luce model
\citep{Turner2019b}, which generalizes Bradley-Terry models to analysis of rankings.
In a Plackett-Luce model, it would typically be the case that tied ``winners'' can occur at any stage of the sequence of choices that forms a multi-item ranking; and this flexibility is what is implemented in the \textbf{PlackettLuce} package. 

The \textbf{PlackettLuce} package also implements a prior penalty for Plackett-Luce models, which regularizes the likelihood with the aim of improving estimation. In particular, use of that prior penalty ensures that the conditions of \citet{Ford1957}, which ensure existence and finiteness of parameter estimates, are always satisfied.
The prior penalty, as implemented in the \textbf{PlackettLuce} package, requires no modification at all to work with the Davidson-Luce model. For full details on the \textbf{PlackettLuce} package and its use, see \citet{Turner2019b}.

\bibliographystyle{chicago}
\bibliography{../bib-papers,../bib-packages}

\appendix

\section{Appendix: Computation via Poisson log-linear model representation} 

Here we use a small, artificial example to show details of
implementation of the Davidson-Luce model in \emph{R}, using maximum
likelihood via a log-linear representation as suggested by
\citet{Fienberg1979}.

\subsection{Davidson-Luce model for a small, contrived example}

We imagine here a 4-player round-robin tournament in which each
`contest' involves exactly 3 of the 4 players. A single round-robin
tournament thus has 4 contests, in this setting.

The data we will use are as follows:

\begin{verbatim}
triples_round_robin <- matrix(c(
    NA, 1, 0, 0,
    1, NA, 1, 0,
    0,  1, NA, 1,
    1, 1, 1, NA),
    4, 4, byrow = TRUE,
    dimnames = list(contest = c("BCD", "ACD", "ABD", "ABC"),
                    winner = c("A", "B", "C", "D"))
    )
triples_round_robin
\end{verbatim}

\begin{verbatim}
##        winner
## contest  A  B  C  D
##     BCD NA  1  0  0
##     ACD  1 NA  1  0
##     ABD  0  1 NA  1
##     ABC  1  1  1 NA
\end{verbatim}

The first contest is won outright by player \(B\); the second is tied
between \(A\) and \(C\); the third is tied between \(B\) and \(D\); and
the fourth is a 3-way tie between \(A\), \(B\) and \(C\).

The simple tournament-scoring system described in Section
\ref{three_items}, with 6 points shared across the winners of each
contest, gives points totals as follows:

\begin{verbatim}
6 * colSums(triples_round_robin / rowSums(triples_round_robin, na.rm = TRUE),
            na.rm = TRUE)
\end{verbatim}

\begin{verbatim}
##  A  B  C  D 
##  5 11  5  3
\end{verbatim}

So in this small tournament \(B\) is the clear winner, with \(A\) and
\(C\) jointly second.

To fit the Davidson-Luce model via its Poisson log-linear
representation, we first expand the data to a form that has a separate
row for each possible outcome of every contest. To do this we will use a
special-purpose function named \texttt{expand\_outcomes} (whose
definition is shown at the end, below).

\begin{verbatim}
expanded_data <- expand_outcomes(triples_round_robin)
print(expanded_data, digits = 2)
\end{verbatim}

\begin{verbatim}
##          comparison    A    B    C    D delta2 delta3 outcome
## 1: B              1 0.00 1.00 0.00 0.00      0      0       1
## 1: C              1 0.00 0.00 1.00 0.00      0      0       0
## 1: D              1 0.00 0.00 0.00 1.00      0      0       0
## 1: B=C            1 0.00 0.50 0.50 0.00      1      0       0
## 1: B=D            1 0.00 0.50 0.00 0.50      1      0       0
## 1: C=D            1 0.00 0.00 0.50 0.50      1      0       0
## 1: B=C=D          1 0.00 0.33 0.33 0.33      0      1       0
## 2: A              2 1.00 0.00 0.00 0.00      0      0       0
## 2: C              2 0.00 0.00 1.00 0.00      0      0       0
## 2: D              2 0.00 0.00 0.00 1.00      0      0       0
## 2: A=C            2 0.50 0.00 0.50 0.00      1      0       1
## 2: A=D            2 0.50 0.00 0.00 0.50      1      0       0
## 2: C=D            2 0.00 0.00 0.50 0.50      1      0       0
## 2: A=C=D          2 0.33 0.00 0.33 0.33      0      1       0
## 3: A              3 1.00 0.00 0.00 0.00      0      0       0
## 3: B              3 0.00 1.00 0.00 0.00      0      0       0
## 3: D              3 0.00 0.00 0.00 1.00      0      0       0
## 3: A=B            3 0.50 0.50 0.00 0.00      1      0       0
## 3: A=D            3 0.50 0.00 0.00 0.50      1      0       0
## 3: B=D            3 0.00 0.50 0.00 0.50      1      0       1
## 3: A=B=D          3 0.33 0.33 0.00 0.33      0      1       0
## 4: A              4 1.00 0.00 0.00 0.00      0      0       0
## 4: B              4 0.00 1.00 0.00 0.00      0      0       0
## 4: C              4 0.00 0.00 1.00 0.00      0      0       0
## 4: A=B            4 0.50 0.50 0.00 0.00      1      0       0
## 4: A=C            4 0.50 0.00 0.50 0.00      1      0       0
## 4: B=C            4 0.00 0.50 0.50 0.00      1      0       0
## 4: A=B=C          4 0.33 0.33 0.33 0.00      0      1       1
\end{verbatim}

The \texttt{expanded\_data} object is an ordinary data frame that can be
used with \emph{R}'s standard functions for fitting generalized linear
models. The Davidson-Luce model could now just be fitted by maximum
likelihood in \emph{R} through a call to \texttt{glm()}, as a Poisson
log-linear model as follows:

\begin{verbatim}
DLmodel <- glm(outcome ~ comparison + A + B + C + D + delta2 + delta3,
               family = poisson, data = expanded_data)
\end{verbatim}

But here the factor named \texttt{comparison} is included purely for
technical reasons, to ensure that the fitted probabilities (over the 7
possible outcomes in each contest here) sum to 1. That factor is not of
any interest, and so for tidiness --- as well as a slight improvement in
computational efficiency --- we will use \texttt{gnm} (from the
\textbf{gnm} package) instead of \texttt{glm}. The advantage of
\texttt{gnm} here is that it allows the `nuisance' factor
\texttt{comparison} to be included more cleanly in the model via the
\texttt{eliminate} argument:

\begin{verbatim}
library(gnm)
DLmodel <- gnm(outcome ~ A + B + C + D + delta2 + delta3, eliminate = comparison,
               family = poisson, data = expanded_data)
DLmodel
\end{verbatim}

\begin{verbatim}
## 
## Call:
## 
## gnm(formula = outcome ~ A + B + C + D + delta2 + delta3, eliminate = comparison, 
##     family = poisson, data = expanded_data)
## 
## Coefficients of interest:
##      A       B       C       D  delta2  delta3  
##  2.071   6.864   2.071      NA   2.390   3.249  
## 
## Deviance:            11.35986 
## Pearson chi-squared: 14.20569 
## Residual df:         19
\end{verbatim}

The reported model parameters are on the log scale; and the
parameterization here has \(\alpha_D\) arbitrarily set to 1, to resolve
parameter redundancy.

So, for example \(\alpha_C/\alpha_D\) is estimated to be
\(\exp(2.07)/1 = 7.93\).

The two tie-prevalence parameters here are both estimated to be very
large: \(\hat\delta_2 = \exp(2.39) = 10.91\) and
\(\hat\delta_3 = \exp(3.25) = 25.8\). This is due to the deliberately
common occurrence of ties in this dataset, in order to demonstrate how
ties are handled; and also the fact that the estimated player strengths
here differ widely. (The data seen here would suggest that in notional
contests where players all have \emph{equal} strengths, ties would be
\emph{extremely} common.)

\subsection{Agreement with full round-robin `points totals'}

Since this was a fully balanced round robin tournament design, then as
mentioned in Section \ref{three_items} the fit of the Davidson-Luce
model should agree exactly with the simple points totals that were
calculated above. Those points totals do indeed agree with their
expectations under the fitted Davidson-Luce model:

\begin{verbatim}
DLfitted <- predict(DLmodel, type = "response")
print(DLfitted, digits = 2)
\end{verbatim}

\begin{verbatim}
##     1: B     1: C     1: D   1: B=C   1: B=D   1: C=D 1: B=C=D     2: A 
##  0.34278  0.00284  0.00036  0.34071  0.12096  0.01101  0.18133  0.02967 
##     2: C     2: D   2: A=C   2: A=D   2: C=D 2: A=C=D     3: A     3: B 
##  0.02967  0.00374  0.32385  0.11498  0.11498  0.38312  0.00284  0.34278 
##     3: D   3: A=B   3: A=D   3: B=D 3: A=B=D     4: A     4: B     4: C 
##  0.00036  0.34071  0.01101  0.12096  0.18133  0.00200  0.24096  0.00200 
##   4: A=B   4: A=C   4: B=C 4: A=B=C 
##  0.23950  0.02181  0.23950  0.25423
\end{verbatim}

\begin{verbatim}
expected_points_totals <- 6 * colSums(expanded_data[, c("A","B","C","D")] * DLfitted)
expected_points_totals
\end{verbatim}

\begin{verbatim}
##         A         B         C         D 
##  5.000000 11.000000  5.000000  3.000001
\end{verbatim}

The actual points totals, from above, were 5, 11, 5 and 3. The agreement
is exact, apart from numerical error due to the iteration-stopping rule
that was used by \texttt{gnm}.

\subsection{Illustration of tie-prevalence interpretations}

The interpretation of tie-prevalence parameters \(\delta_2\) and
\(\delta_3\) was described in Section \ref{three_items}, in terms of the
probabilities in a notional contest involving only players of equal
ability.

Merely as a numerical illustration of those interpretations, we re-fit
here the Davidson-Luce model, but with the constraint that strengths
\(\alpha_A,\alpha_B,\alpha_C,\alpha_D\) are all equal to 1 (so that
their logarithms are all zero).

\begin{verbatim}
DL_equal_strengths <- update(DLmodel, . ~ . - A - B - C - D)
DL_equal_strengths
\end{verbatim}

\begin{verbatim}
## 
## Call:
## gnm(formula = outcome ~ delta2 + delta3, eliminate = comparison, 
##     family = poisson, data = expanded_data)
## 
## Coefficients of interest:
## delta2  delta3  
## 0.6931  1.0986  
## 
## Deviance:            14.90944 
## Pearson chi-squared: 24 
## Residual df:         22
\end{verbatim}

The tie-prevalence estimates here are \(\hat\delta_2 = \exp(0.6931)\)
and \(\hat\delta_3 = \exp(1.0986)\). Agreement with the detailed
interpretations shown in Section \ref{three_items} can thus be checked
as follows:

\begin{verbatim}
coefs <- coef(DL_equal_strengths)
print(round(coefs, 4))
\end{verbatim}

\begin{verbatim}
## Coefficients of interest:
## delta2 delta3 
## 0.6931 1.0986
\end{verbatim}

\begin{verbatim}
delta2 <- exp(coefs[1])
delta3 <- exp(coefs[2])
delta2/(1 + delta2)
\end{verbatim}

\begin{verbatim}
##    delta2 
## 0.6666667
\end{verbatim}

\begin{verbatim}
delta3/(3 + 3*delta2 + delta3)
\end{verbatim}

\begin{verbatim}
## delta3 
##   0.25
\end{verbatim}

These values agree with what was seen in the data, which was 2 two-way
ties out of the 3 contests whose outcome was not a 3-way tie (so
\(\hat\delta_2/(1 + \hat\delta_2) = 2/3\)), and one 3-way tie out of the
4 contests observed in total (so
\(\hat\delta_3/(3 + 3\hat\delta_2 + \hat\delta_3) = 1/4\)).

\subsection{Definition of the function used to expand the data}

For completeness here, we show the full definition of the function that
made the dataframe named \texttt{expanded\_data} in the above.

The function shown here is very much a prototype, not programmed for
efficiency, robustness or scalability.

\begin{verbatim}
expand_outcomes
\end{verbatim}

\begin{verbatim}
## function(m) {
##     n_comparisons <- nrow(m)
##     n_items <- ncol(m)
##     items <- colnames(m)
##     rvec <- apply(m, 1, function(row) sum(!is.na(row)))
##     tvec <- apply(m, 1, function(row) sum(na.omit(row)))
##     maxt <- max(tvec)
##     if (maxt > 1) delta_names <- paste0("delta", 2:maxt)
##     n_possible_outcomes  <- integer(n_comparisons)
##     for (i in 1:n_comparisons) {
##         n_possible_outcomes[i] <- sum(choose(rvec[i], 1:(min(rvec[i], maxt))))
##     }
##     result <- matrix(0, sum(n_possible_outcomes), n_items + maxt + 1)
##     colnames(result) <- c("comparison", colnames(m), delta_names, "outcome")
##     rownames(result) <- as.character(1:nrow(result))
##     filled <- 0
##     for (comparison in 1:n_comparisons){
##         involved <- items[!is.na(m[comparison, ])]
##         for (t in 1:maxt) {
##             combs <- combn(involved, t)
##             for (index in 1:ncol(combs)){
##                 result[filled + index, 1] <- comparison
##                 result[filled + index, 1 + which(items %in% combs[, index])] <- 1/t
##                 if (t > 1) {
##                     result[filled + index, n_items + t] <- 1
##                 }
##                 if (all(na.omit(t * result[filled + index, 1 + (1:n_items)] -
##                                 m[comparison, ]) == 0)) {
##                     result[filled + index, "outcome"] <- 1
##                 }
##                 rownames(result)[filled + index] <-
##                     paste(comparison, paste0(combs[, index], collapse = "="),
##                           sep = ": ")
##             }
##             filled <- filled + ncol(combs)
##         }
##     }
##     result <- as.data.frame(result)
##     result$comparison <- as.factor(result$comparison)
##     return(result)
## }
## <bytecode: 0x36ad6c0>
\end{verbatim}

\end{document}